\documentclass[aps,pra,twocolumn,amsmath,superscriptaddress,longbibliography]{revtex4-2}
\usepackage{amsmath}
\usepackage[urlcolor=blue,colorlinks=true,citecolor=blue,linkcolor=blue,pdfstartview={FitH},bookmarks=false]{hyperref}
\usepackage{appendix}
\usepackage{graphicx}
\usepackage{longtable}
\usepackage{epsfig}
\usepackage{dcolumn}
\usepackage{bm}
\usepackage{bbm}
\usepackage{amssymb}
\usepackage{multirow}
\usepackage{times,color}
\usepackage{hyperref}
\usepackage{amsmath}
\usepackage{color}
\usepackage{subfigure}
\usepackage{float}
\usepackage{epstopdf}
\usepackage{makecell}
\usepackage[normalem]{ulem}

\usepackage[colorlinks=true,
            linkcolor=blue,
            citecolor=blue,
            urlcolor=blue]{hyperref}

\begin{document}

\title{Arithmetic Tuning of Dynamical Critical Exponents in Quasiperiodic Localization Transitions}

\author {Tian-Cheng Yi}
\email{yitiancheng$\textunderscore$phys@163.com}

\affiliation{Department of Physics, Zhejiang Sci-Tech University, Hangzhou 310018, China}
\affiliation{Zhejiang Key Laboratory of Quantum State Control and Optical Field Manipulation,
Zhejiang Sci-Tech University, Hangzhou 310018, China}

\author {Yi-Fan Liu}

\affiliation{Department of Physics, Zhejiang Sci-Tech University, Hangzhou 310018, China}
\affiliation{Zhejiang Key Laboratory of Quantum State Control and Optical Field Manipulation,
Zhejiang Sci-Tech University, Hangzhou 310018, China}

\author {Enguo Guan}

\email{guanenguow@gmail.com}
\affiliation{Innovation Academy for Precision Measurement Science and Technology, Chinese Academy of Sciences, Wuhan 430071, China}

\author{Wen-Long You}
\email{wlyou@nuaa.edu.cn}
\affiliation{College of Physics, Nanjing University of Aeronautics and Astronautics,  Nanjing 211106, China}
\affiliation{Center for the Cross-disciplinary Research of Space Science and Quantum-technologies (CROSS-Q), Nanjing University of Aeronautics and Astronautics,  Nanjing 211106, China}
\begin{abstract}

The critical exponents and universality classes of localization transitions in quasiperiodic systems are of fundamental importance for understanding critical phenomena in aperiodic systems.
Here we show that the dynamical critical behavior can be tuned without adding new terms or changing the form of the Hamiltonian, but solely by varying the incommensurate frequency of the quasiperiodic onsite potential. We construct a family of incommensurate frequencies from the limiting ratios of generalized Fibonacci sequences controlled by the parameters $(m,n)$, and use them to define the quasiperiodic onsite potential. By combining generalized fidelity susceptibility, localization-length scaling, and finite-size gap analysis, we find that the correlation-length exponent is insensitive to the choice of the incommensurate frequency and remains consistent with the correlation-length critical exponent, $\nu \simeq 1$, in the localization transition of the standard Aubry--Andr'e--Harper model. In contrast, the dynamical exponent extracted from the low-energy gap scaling varies systematically with the incommensurate frequency. Our results show that changing the incommensurate frequency provides a simple way to tune dynamical critical scaling in deterministic aperiodic systems. 
Our results suggest instead that the arithmetic structure of an irrational number can serve as a control parameter for nonequilibrium quantum dynamics, enabling the tuning of dynamical critical behavior without changing the microscopic Hamiltonian or the physical spatial dimension.

\end{abstract}
\maketitle

\section{Introduction}

Quantum criticality provides a fundamental framework for understanding collective behavior in low-dimensional many-body systems, where universal scaling laws are governed by a small set of critical exponents~\cite{sachdev2011quantum, fisher1974the, vojta2003quantum}. 
Among the various platforms for exploring such phenomena, quasiperiodic systems occupy a particularly interesting position. 
Unlike disordered systems, they exhibit localization transitions driven by deterministic incommensurate modulations rather than randomness~\cite{aubry1980analyticity, sokoloff1985unusual, kohmoto1983metal}. 
The Aubry--Andr\'e--Harper (AAH) model serves as a canonical model in this context, displaying a localization--delocalization transition at a finite modulation strength~\cite{aubry1980analyticity, harper1955single} and providing a paradigmatic example of criticality in quasiperiodic lattices~\cite{jitomirskaya1999metal, roati2008anderson, lahini2009observation, modugno2009exponential, schreiber2015observation, luschen2018single}. 
Over the past several decades, the AAH model has been extensively investigated and generalized along multiple directions, including long-range hopping, slowly varying or mosaic quasiperiodic potentials, $p$-wave pairing, and non-Hermitian extensions~\cite{biddle2009localization,biddle2010predicted,biddle2011localization,lang2012edge,liu2015localization,ganeshan2015nearest,wang2016phase,cestari2016fate,li2016quantum,longhi2019topological,xu2019butterfly,goblot2020emergence,wang2020realization,wang2020one,cai2021boundary,wang2021many,ahmed2022flat,guan2023reentrant,ren2024identifying,wei2025reentrant}. 
These developments have broadened the scope of the model and established it as a useful platform for exploring localization, topology, and critical phenomena in quasiperiodic systems.

In the standard AAH model, the localization transition is characterized by the correlation-length exponent $\nu=1$. 
Previous studies have shown that modifying the Hamiltonian, for example by introducing disorder, superconducting pairing, external fields, complex onsite potentials, or nonreciprocal hopping, can alter the critical behavior and give rise to distinct scaling behavior~\cite{lv2022quantum,lv2022exploring,liu2024quantum,yi2025unveiling,bu2022quantum,bu2023kibble,liang2024quantum,sahoo2025stark,sun2025non,PhysRevB.101.174203,pmrb-qk7j}. 
However, these mechanisms rely on explicit changes to the model itself. 
A natural but less explored question is whether certain aspects of critical scaling can be tuned while keeping the microscopic form of the Hamiltonian unchanged. 

One possible route is to vary the irrational number that defines the quasiperiodic modulation. 
In most studies, the modulation frequency is chosen as the inverse golden ratio $(\sqrt{5}-1)/2$, without explicitly considering the possible impact of alternative irrational numbers~\cite{biddle2009localization,biddle2010predicted,biddle2011localization,lang2012edge,kraus2012topological,liu2015localization,ganeshan2015nearest,wang2016phase,cestari2016fate}. 
However, from a number-theoretic perspective, irrational numbers can be associated with distinct continued-fraction expansions, recursive constructions, and substitution rules~\cite{avila2015global, damanik2017schrodinger, last1996quantum}, leading to different hierarchical structures in real space. 
These structures provide a possible route to affect quasiperiodic critical scaling without introducing additional terms into the Hamiltonian.

To explore this possibility, we employ a class of quasiperiodic modulations generated by the generalized Fibonacci recursion
$F_{k+1}=nF_k+mF_{k-1}$, whose associated irrational number is
$\Omega_{n,m}=(\sqrt{n^2+4m}-n)/(2m)$~\cite{macia2017spectral}.
This construction defines a systematic arithmetic family that extends the conventional Fibonacci sequence and provides a concrete way to vary the irrational number. From a number-theoretic point of view, different choices of $(m,n)$ determine distinct asymptotic structures of the continued-fraction approximants to $\Omega_{n,m}$. In the lattice problem, these approximants set the hierarchy of length scales over which the quasiperiodic potential nearly repeats itself. Therefore, varying $(m,n)$ changes the multiscale pattern of quasi-resonances experienced by the wave functions, while leaving the microscopic Hamiltonian unchanged. This provides a controllable framework for examining whether the arithmetic properties of the quasiperiodic modulation can affect the scaling behavior of quasiperiodic localization transitions.

In this work, we systematically investigate the localization transition in the AAH model with generalized Fibonacci irrational modulations by means of fidelity susceptibility, localization length, and finite-size gap scaling. 
Our main results show that, for different values of $m$ and $n$, the correlation-length exponent $\nu$ remains consistent with the value $\nu=1$ in the standard AAH model. 
This conclusion is supported independently by finite-size scaling analyses of both the fidelity susceptibility and the localization length. 
In contrast, the dynamical critical exponent $z$, extracted from the critical gap scaling $\Delta(V_c)\sim N^{-z}$, depends on the arithmetic parameters $m$ and $n$. 
These results indicate that the correlation-length exponent remains AAH-like, whereas the dynamical critical exponent can be tuned solely by changing the arithmetic structure of the irrational modulation. 
Thus, the irrational number is not merely a parameter used to generate quasiperiodicity; it also serves as an additional control parameter for dynamical critical scaling.

The organization of the paper is as follows. 
In Sec.~\ref{sec2}, we introduce the AAH model with generalized Fibonacci irrational modulations. 
In Sec.~\ref{sec3}, we define the main physical quantities used in this work, including the fidelity susceptibility, localization length, and excitation gap, together with their finite-size scaling forms. 
In Sec.~\ref{sec4}, we present the numerical results for the localization transition and extract the correlation-length exponent $\nu$ and the dynamical exponent $z$ for different arithmetic classes of irrational modulations. 
Finally, Sec.~\ref{sec5} summarizes our results.

\section{Model}
\label{sec2}
We consider a one-dimensional tight-binding chain of spinless fermions subject to a quasiperiodic onsite potential,
\begin{equation}
H = \sum_{j=1}^{N} \Big[
- J \left(c_j^\dagger c_{j+1} + c_{j+1}^\dagger c_j\right)
+ V_j\, c_j^\dagger c_j
\Big],
\label{ham}
\end{equation}
where $c_j^\dagger$ ($c_j$) creates (annihilates) a fermion on site $j$, and $J>0$ denotes the nearest-neighbor hopping amplitude. The onsite quasiperiodic potential is chosen as $V_j = V \cos(2\pi \Omega j+\phi)$, where $V$ denotes the modulation amplitude, $\phi$ is a phase offset, and $\Omega$ is an irrational number specifying the incommensurate modulation.

For a single-particle eigenstate, we denote by $\psi_j$ the wave-function amplitude on the $j$th lattice site.
The single-particle Schrödinger equation reads 
\begin{eqnarray}
    J(\psi_{j+1}+\psi_{j-1})+V\cos(2\pi\Omega j+\phi)\psi_j=E\psi_j. 
\end{eqnarray}
We introduce the dual-space representation
$\psi_j=\sum_k f_k e^{ik(2\pi\Omega j+\phi)}$, 
where the phase $\phi$ of the quasiperiodic potential has been absorbed into the dual basis. 
Substituting the expansion into the Schrödinger equation gives 
 \begin{equation}
V(f_{k+1}+f_{k-1})+4J\cos(2\pi\Omega k)f_k
=
2Ef_k.
 \label{eq:dual_equation}
 \end{equation}
The dual transformation maps the ratio $V/J$ to $4J/V$. Imposing self-duality gives
$V/J=4J/V$, and hence $V_c=2J$. Throughout this paper, we set $J=1$ as the unit of energy. 
 
At the critical point $V=2$, the Hamiltonian is invariant under this dual transformation, signaling a self-dual point. As a direct consequence of this self-duality, all single-particle eigenstates undergo a simultaneous localization--delocalization transition at $V=2$, thereby excluding the possibility of a mobility edge.

To investigate the role of arithmetic structure in a systematic way, we generate $\Omega$ from a generalized Fibonacci sequence, $F_1 = 1$, $F_2 = m+n$, $F_{k+1} = mF_{k-1}+nF_k$, $ n,m \in \mathbb{Z}^+$, whose ratio of successive terms converges to $\Omega_{n,m} = \lim_{k\to\infty} \frac{F_k}{F_{k+1}} = \frac{\sqrt{n^2+4m}-n}{2m}$.
This construction generates a two-parameter family of irrational modulations that extends the conventional noble-metal sequence. In finite-size calculations, we approximate $\Omega_{n,m}$ by the rational convergent $F_k/F_{k+1}$ and choose the system size as $N=F_{k+1}$, which provides a controlled approach to the quasiperiodic limit.
The representative binary patterns generated by the substitution rule are discussed in Appendix~\ref{app:substitution_sequences}.

A subtle but important point is that, for certain choices of $(m,n)$, two successive generalized Fibonacci numbers are not coprime. For example, for $(m,n)=(2,2)$, the sequence reads $1,\;4,\;10,\;28,\;76,\;208,\;568,\;1552,\;4240,\;11584,\ldots$, 
so that the rational approximant $F_k/F_{k+1}$ may contain a common divisor. In such cases, we first reduce the fraction $F_k/F_{k+1}=p_k/q_k$ to its coprime form and then choose the system size as $N=q_k$. This prescription removes artificial commensurability effects caused by unreduced approximants and avoids spurious estimates of critical exponents.

We also exclude parameter choices for which $\Omega_{n,m}$ becomes rational. Since $n^2+4m$ is an integer, $\Omega_{n,m}$ is irrational only when $n^2+4m$ is not a perfect square. For instance, $(m,n)=(2,1)$ gives $\Omega_{2,1}=1/2$, while $(m,n)=(3,2)$ gives $\Omega_{3,2}=1/3$. These rational cases correspond to periodic rather than quasiperiodic modulations and are therefore not considered in the following analysis.

\section{Physical quantities}
\label{sec3}
\subsection{Generalized fidelity susceptibility}

Fidelity susceptibility has become a widely used and conceptually transparent diagnostic of quantum criticality from the geometry of quantum states ~\cite{PhysRevE.74.031123,PhysRevE.76.022101,PhysRevLett.99.095701, PhysRevLett.99.100603,gu2010fidelity}. As the leading quadratic response of the quantum fidelity to an infinitesimal change of a control parameter, it directly measures the sensitivity of the many-body wave function and is closely related to the quantum geometric tensor and dynamical response functions~\cite{PhysRevE.76.022101,PhysRevLett.99.095701,PhysRevLett.99.100603}. Owing to its order-parameter-free character, this approach has been successfully applied to conventional symmetry-breaking transitions, topological phase transitions, finite-temperature quantum critical regimes, strongly correlated lattice models, matrix-product-state descriptions, and large-scale numerical simulations based on quantum Monte Carlo methods ~\cite{PhysRevA.75.032109,PhysRevA.76.062318,PhysRevA.78.010301,PhysRevA.79.032302,PhysRevB.76.104420,PhysRevLett.103.170501,PhysRevB.81.064418,PhysRevX.5.031007}. These developments demonstrate that fidelity susceptibility is not only a sensitive locator of critical points, but also a useful scaling quantity for extracting universal critical information~\cite{PhysRevB.77.245109,gu2009scaling,PhysRevLett.106.055701}. Such advantages are particularly important in disordered and quasiperiodic systems, where conventional symmetry-breaking order parameters may be absent or difficult to identify.
In the context of localization transitions, fidelity-based quantities provide a natural probe of the critical response of eigenstates to variations in the quasiperiodic potential strength. Although fidelity susceptibility has been widely used in various quantum critical systems, the finite-size scaling properties of its generalized higher-order forms in quasiperiodic localization transitions have not been systematically characterized. In what follows, we investigate the fidelity susceptibility and its generalized counterparts in the generalized AAH model, with particular emphasis on their finite-size scaling behavior.

For a parameter-dependent Hamiltonian $\hat{H}(\lambda)$, the generalized
fidelity susceptibility (GFS) associated with the eigenstate
$|\psi_n(\lambda)\rangle$ is defined as ~\cite{you2015generalized} 
\begin{equation}
\chi^{(n)}_{2r+2}(\lambda)
=
\sum_{m\neq n}
\frac{
\left|
\left\langle \psi_m(\lambda)\right|
\partial_{\lambda}\hat{H}
\left|\psi_n(\lambda)\right\rangle
\right|^2
}
{
\left[
E_m(\lambda)-E_n(\lambda)
\right]^{2r+2}
},
\label{eq:gfs_def}
\end{equation}
where $E_n(\lambda)$ is the corresponding eigenenergy and
$\partial_\lambda\hat{H}$ denotes the perturbation operator generated by the
driving parameter $\lambda$. 
The conventional fidelity susceptibility is recovered from Eq.~\eqref{eq:gfs_def}
by taking $r=0$, namely~\cite{PhysRevE.76.022101}
\begin{equation}
\chi^{(n)}_2(\lambda)
=
\sum_{m\neq n}
\frac{
\left|
\left\langle \psi_m(\lambda)\right|
\partial_{\lambda}\hat{H}
\left|\psi_n(\lambda)\right\rangle
\right|^2
}
{
\left[
E_m(\lambda)-E_n(\lambda)
\right]^2
}.
\label{eq:chi2_spectral}
\end{equation}
This quantity can also be interpreted geometrically as the Riemannian metric
tensor on the parameter manifold~\cite{provost1980riemannian,gu2010fidelity}. 
For a single driving parameter, the diagonal component of the quantum metric
tensor is given by
$g^{(n)}_{\lambda\lambda}
=
\mathrm{Re}
\left[
\langle \partial_\lambda \psi_n|\partial_\lambda \psi_n\rangle
-
\langle \partial_\lambda \psi_n|\psi_n\rangle
\langle \psi_n|\partial_\lambda \psi_n\rangle
\right]$,
which is equivalent to the spectral expression,
i.e.,
$g^{(n)}_{\lambda\lambda}
=
\chi^{(n)}_2(\lambda)$.
More generally, the quantum geometric tensor is a complex tensor whose real
part gives the quantum metric, while its imaginary antisymmetric part encodes
the Berry curvature and hence the geometric Berry phase accumulated along a
closed path in parameter space~\cite{berry1984quantal,provost1980riemannian}.
For pure states, the corresponding quantum Fisher information is simply four
times the quantum metric, or equivalently
$F_Q^{(n)}=4g^{(n)}_{\lambda\lambda}=4\chi^{(n)}_2$ for a single parameter
estimation problem~\cite{braunstein1994statistical,liu2014fidelity}.
Thus, $\chi_2$ measures the intrinsic distance between neighboring quantum
states and provides a geometric characterization of the critical deformation
of the wave function.
In the present work, the quasiperiodic potential strength is chosen as the
driving parameter, $\lambda=V$. 
Since the generalized AAH model considered here does not exhibit a
mobility edge, the localization transition is not energy-selective in the
spectrum. 
Therefore, in the following we focus on the GFS of the lowest eigenstate
$|\psi_1(V)\rangle$. 
For notational simplicity, the superscript of
$\chi^{(1)}_{2r+2}$ is omitted hereafter, and we write
$\chi_{2r+2}\equiv\chi^{(1)}_{2r+2}$ for
$r=0,1$. 
In particular, $\chi_2$ denotes the conventional fidelity susceptibility,
whereas $\chi_4$ denotes the fourth-order generalized fidelity susceptibility.
Compared with $\chi_2$, the higher-order quantity $\chi_4$ contains a larger
power of the excitation energy in the denominator and is therefore more
sensitive to the low-energy critical modes near the localization transition.

Near the critical point, the GFS is expected to obey the finite-size scaling
form~\cite{PhysRevB.77.245109,gu2009scaling,you2015generalized}
\begin{equation}
\chi_{2r+2}(V,N)
=
N^{2/\nu+2rz}
\,
\Phi_r
\left[
N^{1/\nu}(V-V_c)
\right],
\label{eq:gfs_scaling}
\end{equation}
where $N$ is the system size, $\nu$ is the correlation-length exponent, $z$ is the dynamical critical exponent, and $\Phi_r$ is a scaling function. 
The scaling form used above follows from the standard finite-size scaling of generalized adiabatic susceptibilities. Near a localization critical point, the relevant scaling variable is $N^{1/\nu}(V-V_c)$, while each additional pair of energy denominators in the generalized susceptibility contributes a factor set by the low-energy scale $\Delta\sim N^{-z}$. This gives the exponent $2/\nu+2rz$ for $\chi_{2r+2}$, consistent with previous fidelity-susceptibility scaling analyses.

Equivalently, if the maximum of $\chi_{2r+2}$ occurs at the finite-size pseudocritical point $V_m$, then
\begin{equation}
\chi_{2r+2,\max}
\equiv
\chi_{2r+2}(V_m,N)
\sim
N^{d_{2r+2}},
\label{eq:gfs_peak_scaling}
\end{equation}
where $d_{2r+2}
=
\frac{2}{\nu}+2rz$.
Once the exponent $\nu$ is determined independently from the scaling collapse of $\chi_2$ or other localization observables, the dynamical exponent can be extracted as
\begin{equation}
z
=
\frac{1}{2r}
\left(
d_{2r+2}-\frac{2}{\nu}
\right).
\label{eq:z_from_gfs}
\end{equation}
For the fourth-order GFS, corresponding to $r=1$, this relation becomes
\begin{equation}
z
=
\frac{1}{2}
\left(
d_4-\frac{2}{\nu}
\right).
\label{eq:z_from_chi4}
\end{equation}
Therefore, $\chi_4$ provides a static ground-state route to the dynamical
critical exponent, complementary to the direct finite-size scaling of the
excitation gap.

This scaling strategy is motivated by previous studies of fidelity
susceptibility and generalized fidelity susceptibility in disordered and 
quasiperiodic systems, where fidelity-based probes were
shown to efficiently identify critical points and extract universal critical
exponents~\cite{lv2022quantum,lv2022exploring,liu2024quantum,yi2025unveiling,ren2024identifying}.
Here we extend this perspective to the generalized Fibonacci quasiperiodic
modulation and use $\chi_2$ and $\chi_4$ to characterize the critical scaling
of the transition in the generalized AAH model. 
\subsection{Localization length}

To characterize the spatial extent of single-particle wave functions, we calculate the localization length.
For a normalized eigenstate
$|\psi\rangle=\sum_{j=1}^{N}\psi_j |j\rangle$,
where $|j\rangle$ denotes the single-particle basis state localized
on site $j$.
The probability distribution on site $j$ is
defined as
$P_j = |\psi_j|^2$,
with $\sum_j P_j=1$. The localization center is given by
$j_c = \sum_j j P_j $.
The localization length is then defined as the root-mean-square spatial spread of the wave function,
\begin{equation}
\xi =
\sqrt{
\sum_j (j-j_c)^2 P_j
}.
\label{eq:loc_length}
\end{equation}
For an extended state, $\xi$ grows proportionally with the system size $N$, whereas for a localized state, $\xi$ remains finite in the thermodynamic limit. Near the localization transition, $\xi$ is expected to diverge as
$\xi \sim |V-V_c|^{-\nu}$,
where $V_c$ is the critical point and $\nu$ is the correlation-length critical exponent. Therefore, the finite-size scaling form can be written as
\begin{equation}
\frac{\xi}{N}
=
f\!\left[(V-V_c)N^{1/\nu}\right],
\label{scaling_xi}
\end{equation}
which allows us to extract $\nu$ from the scaling collapse of numerical data.
In this work, the localization length $\xi$ is always evaluated for the lowest-energy single-particle eigenstate.

\subsection{Energy gap}

In addition to fidelity susceptibility and localization length, we also analyze the low-energy excitation gap. 
For a finite system of size $N$, the energy gap is defined as the difference between the first-excited-state energy and the ground-state energy,
\begin{equation}
\Delta(V)
=
E_1(V,N)-E_0(V,N),
\label{eq:gap_def}
\end{equation}
where $E_0$ and $E_1$ denote the lowest and the first-excited eigenenergies of the Hamiltonian, respectively. 
The closing of this gap provides an independent signature of the critical point.

For a finite system, the corresponding finite-size scaling form is
\begin{equation}
\Delta(V)
=
N^{-z}
\mathcal{G}
\left[
(V-V_c)N^{1/\nu}
\right],
\label{eq:gap_scaling}
\end{equation}
where $\mathcal{G}$ is a scaling function. 
Equivalently, the rescaled quantity $\Delta N^z$ should collapse onto a single universal curve when plotted as a function of $(V-V_c)N^{1/\nu}$. 
At the critical point, Eq.~(\ref{eq:gap_scaling}) reduces to
$\Delta(V_c)
\sim
N^{-z}$.
Therefore, the dynamical exponent $z$ can be extracted from the slope of $\ln \Delta(V_c)$ versus $\ln N$,
$\ln \Delta(V_c)
=
-z\ln N+\mathrm{const}$.

In practical calculations, we also consider the minimum gap $\Delta_{\mathrm{min}}$ in the vicinity of the transition point. 
It obeys the same finite-size scaling behavior,
$\Delta_{\mathrm{min}}
\sim
N^{-z}$.
Thus, the gap scaling provides an independent way to determine the dynamical critical exponent and to test the consistency of the critical scaling obtained from fidelity susceptibility and localization length.
\section{Results for scaling behavior}
\label{sec4}
\subsection{The correlation-length
exponent $\nu$ for $m=1$}
\begin{figure}[htbp]
\centering
\includegraphics[width=1.0\linewidth]{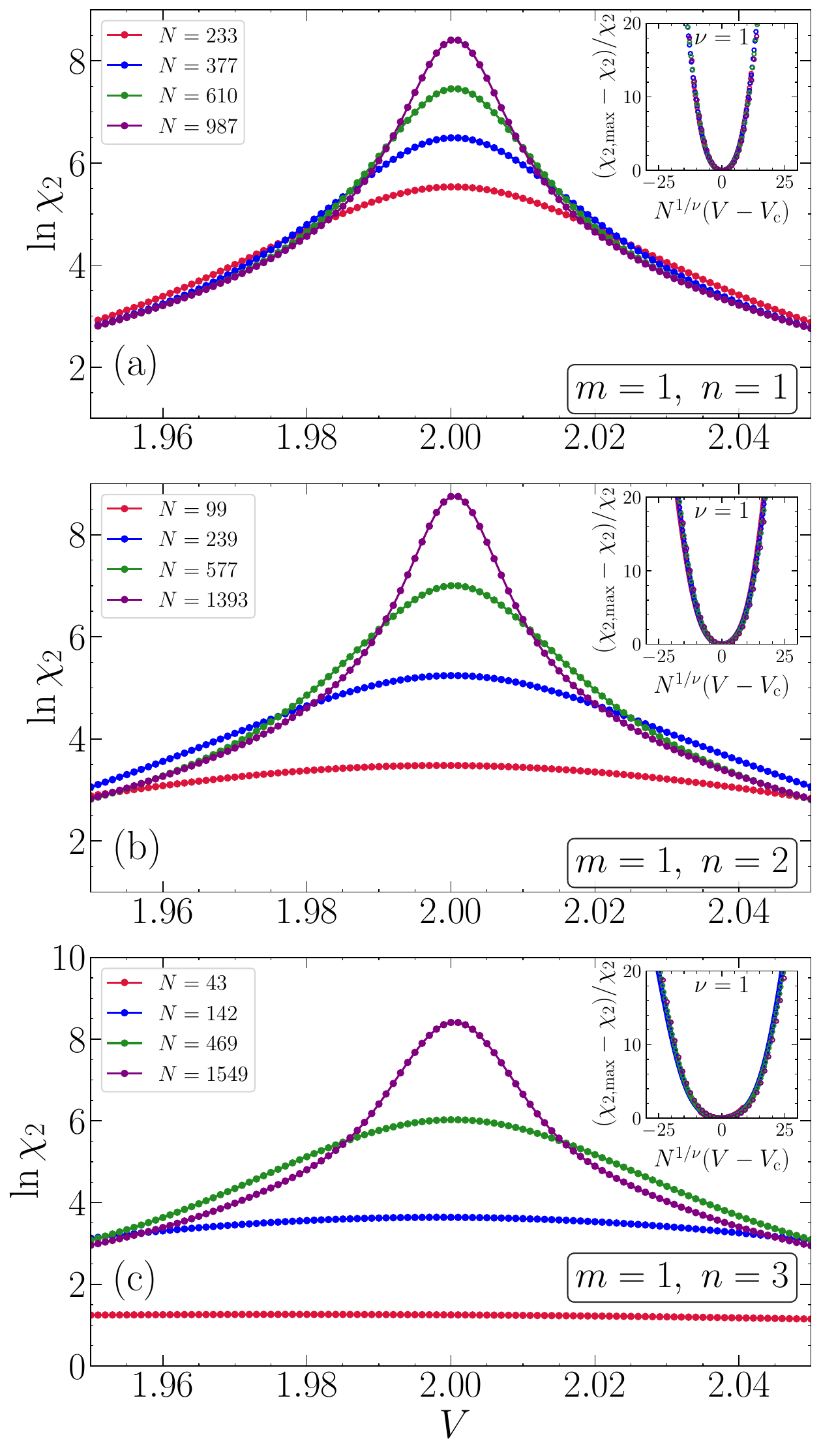}
\caption{
The logarithm of the fidelity susceptibility $\chi_2$
as a function of the quasiperiodic potential strength $V$ for $m=1$:
(a) $n=1$; 
(b) $n=2$; 
(c) $n=3$. 
The phase offset is fixed at $\phi=\pi$. 
The insets show the rescaled fidelity susceptibility $(\chi_{2,\max}-\chi_2)/\chi_2$ as a function of the scaling variable $N^{1/\nu}(V-V_c)$ for $\nu=1$. 
For all three cases, the critical point is $V_c=2$, and the data collapse supports the correlation-length exponent $\nu=1$.
}
\label{fig:FS_m1_n123}
\end{figure}
We begin by analyzing the finite-size scaling behavior of the fidelity susceptibility for the generalized Fibonacci modulations with $m=1$. 
The logarithm of the fidelity susceptibility, $\ln\chi_2$, is plotted in Fig.~\ref{fig:FS_m1_n123} as a function of the quasiperiodic potential strength $V$ for three representative cases, $(m,n)=(1,1)$, $(1,2)$, and $(1,3)$. 
For each modulation, $\ln\chi_2$ exhibits a pronounced peak near the self-dual critical point $V_c=2$. 
With increasing system size, the peak becomes progressively sharper and higher, reflecting the growing sensitivity of the eigenstate to an infinitesimal variation of $V$ in the vicinity of the localization transition.

The insets of Fig.~\ref{fig:FS_m1_n123} show the corresponding finite-size scaling collapse. 
Here we plot the rescaled quantity $(\chi_{2,\max}-\chi_2)/\chi_2$ as a function of the scaling variable $N^{1/\nu}(V-V_c)$, with $V_c=2$ fixed by the self-duality of the model. 
For all three cases, the numerical data for different system sizes collapse well onto a single curve when $\nu=1$ is used. 
This result demonstrates that the correlation-length exponent remains $\nu=1$ for the $m=1$ family, despite the change of the arithmetic structure controlled by $n$. 
Therefore, these generalized Fibonacci modulations share the same correlation-length critical exponent $\nu$ as the conventional Aubry-Andr\'e model.
\begin{figure}[t]
    \centering
    \includegraphics[width=0.5\textwidth]{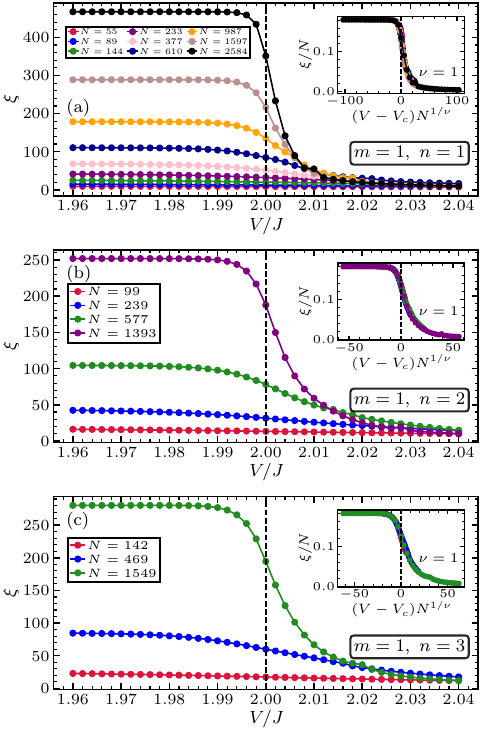}
\caption{
The localization length $\xi$ as a function of the quasiperiodic potential strength $V$ for $m=1$:
(a) $n=1$;
(b) $n=2$;
(c) $n=3$.
The vertical dashed line marks the self-dual critical point $V_c=2$.
The insets show the rescaled localization length $\xi/N$ as a function of the scaling variable $N^{1/\nu}(V-V_c)$ for $\nu=1$. 
All results shown here are averaged over 50 random samples of the phase $\phi$.
}
    \label{fig:xi_collapse_m1_n123}
\end{figure}

To further verify the critical behavior obtained from fidelity susceptibility, 
we also analyze the localization length $\xi$ for the generalized Fibonacci
modulations with $m=1$.  The results are presented in Fig.~\ref{fig:xi_collapse_m1_n123}.
For $(m,n)=(1,1)$, $(1,2)$, and $(1,3)$, the localization length exhibits a
pronounced size-dependent crossover in the vicinity of the self-dual critical
point $V_c/J=2$.  In the extended regime, $\xi$ grows with the system size,
whereas in the localized regime it remains finite.  This behavior is consistent
with the localization transition expected for the generalized AAH model.

Near the critical point, the localization length obeys the finite-size scaling
form (\ref{scaling_xi}). 
As shown in the insets of Fig.~\ref{fig:xi_collapse_m1_n123}, the data for
different system sizes collapse well onto a single curve when $V_c/J=2$ and
$\nu=1$ are used.  The same scaling behavior is observed for all three choices
of $n$, demonstrating that the generalized Fibonacci modulations with $m=1$
share the same correlation-length critical exponent as the conventional
AAH model.
\subsection{The dynamical exponent $z$ for $m=1$}
\begin{figure}[t]
    \centering
    \includegraphics[width=0.95\linewidth]{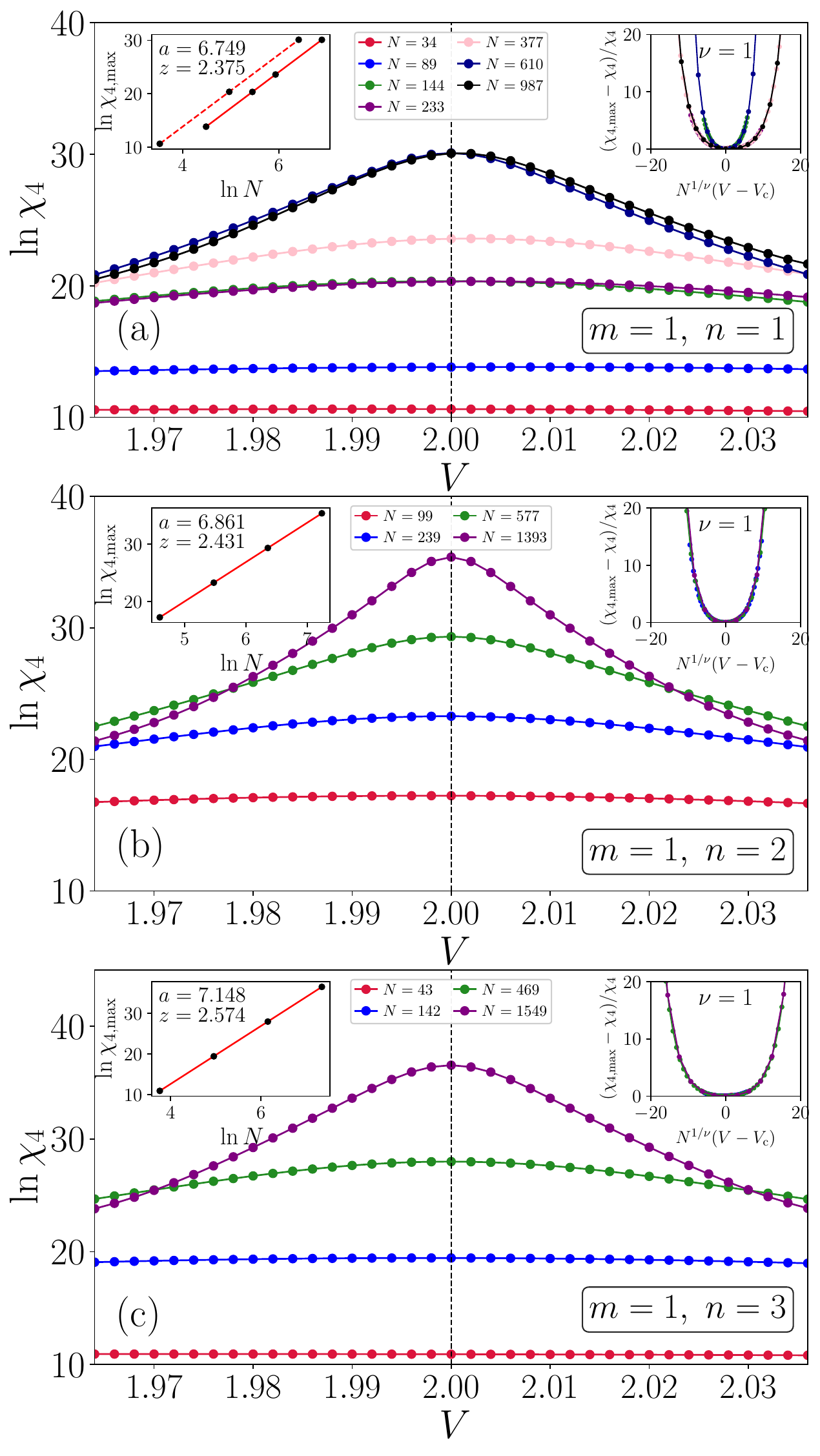}
\caption{
Finite-size scaling of the fourth-order fidelity susceptibility $\chi_4$ for the generalized AAH model with $m=1$ and $n=1,2,3$. The phase offset is fixed at $\phi=\pi$. 
The left insets show the log--log fit of the peak value $\chi_{4,\max}$ according to
$\ln \chi_{4,\max}=a\ln N+b$.
Using the scaling relation $\chi_{4,\max}\sim N^{2/\nu+2z}$ and fixing $\nu=1$, the dynamical exponent is obtained as
$z=(a-2)/2$.
The extracted values are $z=2.375\pm 0.004$, $2.431\pm 0.001$, and $2.574\pm0.002$ for $n=1$, $2$, and $3$, respectively.
The right insets show the corresponding scaling collapse with fixed $\nu=1$.
}
    \label{fig:chi4_m1_n123_z}
\end{figure}

\begin{figure}[t]
    \centering
    \includegraphics[width=0.95\columnwidth]{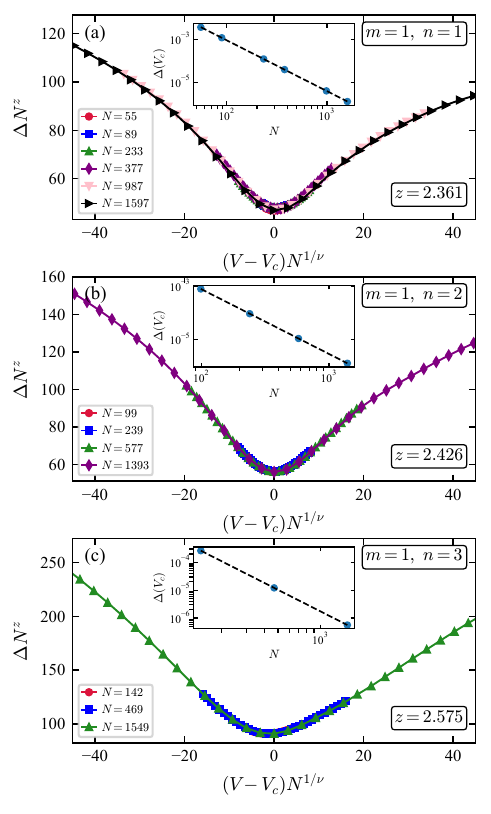}
    \caption{
The rescaled excitation gap ${\Delta}N^z$ as a function of the scaling variable $(V-V_c)N^{1/\nu}$ for $m=1$:
(a) $n=1$;
(b) $n=2$;
(c) $n=3$.
The insets show the log-log scaling of the critical gap ${\Delta}(V_c)$ with the system size $N$.
For all three cases, the critical point is fixed at the self-dual point $V_c=2$, and the data collapse is obtained with $\nu=1$.
The corresponding dynamical exponents are $z=2.361$, $2.426$, and $2.575$ for $n=1$, $2$, and $3$, respectively. All results shown here are averaged over more than 500 random samples of the phase $\phi$.
}
\label{fig:GapCollapse_m1_n123}
\end{figure}

\begin{table}[t]
\caption{
Dynamical critical exponent $z$ for different irrational modulation
frequencies $\Omega_{1,n}$ in the generalized AAH model with $m=1$.
Here $\Omega_{1,n}=(\sqrt{n^2+4}-n)/2$.
}
\label{tab:z_exponents_m1}
\begin{ruledtabular}
\begin{tabular}{cccc}
$(m,n)$ & $\Omega_{m,n}$ & $z_{\chi_4}$ & $z_{\mathrm{gap}}$ \\
\hline
$(1,1)$  & $0.618034$ & $2.375 \pm 0.004$ & $2.361 \pm 0.004$ \\
$(1,2)$  & $0.414214$ & $2.431 \pm 0.001$ & $2.426 \pm 0.001$ \\
$(1,3)$  & $0.302776$ & $2.574 \pm 0.002$ & $2.575 \pm 0.000$ \\
$(1,4)$  & $0.236068$ & $2.805 \pm 0.004$ & $2.804 \pm 0.005$ \\
$(1,5)$  & $0.192582$ & $3.125 \pm 0.017$ & $3.113 \pm 0.002$ \\
$(1,6)$  & $0.162278$ & $3.470 \pm 0.031$ & $3.440 \pm 0.001$ \\
$(1,7)$  & $0.140055$ & $3.890 \pm 0.041$ & $3.776 \pm 0.000$ \\
$(1,8)$  & $0.123106$ & $4.223 \pm 0.067$ & $4.108 \pm 0.000$ \\
$(1,9)$  & $0.109772$ & $4.567 \pm 0.078$ & $4.431 \pm 0.001$ \\
$(1,10)$ & $0.099020$ & $4.895 \pm 0.084$ & $4.748 \pm 0.000$ \\
\end{tabular}
\end{ruledtabular}
\end{table}

\begin{figure}[t]
    \centering
    \includegraphics[width=0.48\textwidth]{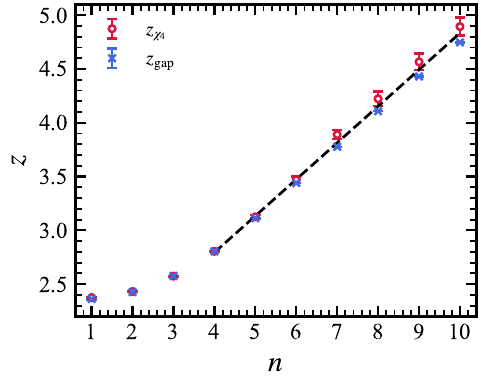}
    \caption{
Dynamical critical exponent $z$ as a function of the integer $n$
in the generalized AAH model at fixed $m=1$.
    The red circles denote the exponent $z_{\chi_4}$ extracted from
    the scaling of the fourth-order fidelity susceptibility, while the
    blue crosses denote $z_{\rm gap}$ extracted from the finite-size
    scaling of the critical gap. Error bars represent the fitting
    uncertainties of the corresponding scaling analyses.
    The black dashed line is an unweighted linear fit to the averaged
    exponent
    $z_{\rm ave}=(z_{\chi_4}+z_{\rm gap})/2$
    using the data with $n\geq 4$.
    The fit gives $z_{\rm ave}=an+b$, with
$a=0.340\pm0.003$ and $b=1.43\pm0.02$.
    The data for $n<4$ are shown for comparison but are not included
    in the linear fit.
    }
    \label{fig:z_vs_n_fit}
\end{figure}

Figure~\ref{fig:chi4_m1_n123_z} presents the finite-size scaling analysis of the fourth-order fidelity susceptibility $\chi_4$ for the generalized AAH model with $m=1$ and $n=1,2,3$.
In all three cases, the peak of $\ln\chi_4$ is centered around the self-dual critical point $V_c=2$, as indicated by the vertical dashed line.
With increasing system size, the peak becomes sharper and its height grows systematically, signaling the enhancement of critical fluctuations near the localization transition.

To extract the dynamical critical exponent, we analyze the system-size dependence of the peak value $\chi_{4,\max}$.
At criticality, the fourth-order fidelity susceptibility satisfies the scaling form $\chi_{4,\max}\sim N^{2/\nu+2z}$.
Therefore, a linear fit in the double-logarithmic scale,
$\ln \chi_{4,\max}=a\ln N+b$,
gives the slope
$a=2/\nu+2z$.
The left inset of each panel shows this log--log fit of
$\ln \chi_{4,\max}$ versus $\ln N$.
The fitted slopes are $a=6.749$, $6.861$, and $7.148$ for
$(m,n)=(1,1)$, $(1,2)$, and $(1,3)$, respectively.
For $(m,n)=(1,1)$, the small oscillation between two branches in the left inset of Fig.~\ref{fig:chi4_m1_n123_z}(a) can be traced to the parity of the finite rational approximants. In this case, the system sizes are Fibonacci numbers $N=F_{k+1}$, with $F_{k+1}=F_k+F_{k-1}$, and their parity repeats with period three: two consecutive Fibonacci sizes are odd and the next one is even, for example $55,89,144,233,377,610,\ldots$. Thus the odd and even Fibonacci approximants fall into two slightly shifted finite-size subsequences. Similar sensitivity to rational-approximant sequences is often encountered in finite-size analyses of quasiperiodic criticality~\cite{lv2022exploring,liu2024quantum}. 
Since the scaling collapse is performed with fixed $\nu=1$, the dynamical
critical exponent is obtained from
$z=(a-2)/2$.
This gives
$z=2.375\pm0.004,\ 2.431\pm0.001,\ 2.574\pm 0.002$
for $n=1$, $2$, and $3$, respectively.
These results show that the dynamical exponent varies with the arithmetic
structure of the quasiperiodic modulation.

The right inset of each panel further shows the finite-size collapse of
$(\chi_{4,\max}-\chi_4)/\chi_4$ as a function of
$N^{1/\nu}(V-V_c)$.
Here we fix the correlation-length exponent to $\nu=1$.
The good data collapse for all three values of $n$ indicates that the
correlation-length exponent remains consistent with the value in the standard AAH model,
whereas the dynamical exponent $z$ is sensitive to the generalized Fibonacci
sequence.
This demonstrates that the generalized quasiperiodic modulation can tune the
dynamical critical behavior while preserving the AA-like exponent $\nu=1$.

The scaling behavior of the excitation gap for the generalized Fibonacci modulations with $m=1$ is shown in Fig.~\ref{fig:GapCollapse_m1_n123}.
According to Eq.~(\ref{eq:gap_scaling}), the rescaled quantity ${\Delta}N^z$ should collapse onto a universal curve when plotted as a function of $(V-V_c)N^{1/\nu}$.
Using the critical point $V_c=2$ and the correlation-length exponent $\nu=1$ obtained from the fidelity-susceptibility analysis, we find a good data collapse for all three representative cases, namely $(m,n)=(1,1)$, $(1,2)$, and $(1,3)$.

The insets display the log-log scaling of the critical gap ${\Delta}(V_c)$ with the system size $N$.
As expected from Eq.~(\ref{eq:gap_scaling}), ${\Delta}(V_c)$ follows a power-law behavior as $N$ increases, from which the dynamical exponent $z$ can be extracted.
The fitted values are $z=2.361\pm0.004$ for $(m,n)=(1,1)$, $z=2.426\pm0.001$ for $(m,n)=(1,2)$, and $z=2.575\pm0.000$ for $(m,n)=(1,3)$.
These results show that, even within the family of generalized Fibonacci modulations with fixed $m=1$, different choices of $n$ lead to different dynamical critical exponents.
Since changing $n$ directly modifies the irrational number $\Omega_{n,1}$, the dynamical exponent $z$ can be tuned solely by changing the arithmetic structure of the quasiperiodic modulation.
Therefore, the arithmetic property of the incommensurate potential provides an additional route to modifying the critical behavior and may drive the system into different universality classes.

The values of $z$ obtained from the gap scaling and from the $\chi_4$ scaling
are in good agreement for all three generalized Fibonacci modulations.
For $(m,n)=(1,1)$ and $(1,2)$, the $\chi_4$ estimates are slightly larger than
the gap-based estimates, while for $(1,3)$ the two results agree within the
quoted error bars.

Table~\ref{tab:z_exponents_m1} and Fig.~\ref{fig:z_vs_n_fit} summarize the dynamical
critical exponent extracted from the fourth-order generalized fidelity
susceptibility and from finite-size gap scaling. 
As shown in Fig.~\ref{fig:z_vs_n_fit}, 
for sufficiently large $n$, the averaged exponent
$z_{\rm ave}=(z_{\chi_4}+z_{\rm gap})/2$ exhibits an approximately
linear dependence on $n$. 
The fit gives $z_{\rm ave}=an+b$, with
$a=0.340\pm0.003$ and $b=1.43\pm0.02$, 
indicating that the dynamical critical exponent can be continuously
enhanced by changing the irrational arithmetic parameter.

These results show that the generalized Fibonacci modulations with $m=1$ share the same correlation-length exponent $\nu=1$, while the dynamical exponent $z$ depends on the arithmetic structure of the incommensurate modulation.
Therefore, the gap scaling is consistent with the fidelity-susceptibility analysis and further confirms that these systems have the same correlation-length critical exponent as the AAH model, although their dynamical critical behavior is not identical.

\subsection{The correlation-length
exponent $\nu$ for $m>1$}

\label{app:xi_scaling}
\begin{figure}[t]
\centering
\includegraphics[width=0.95\linewidth]{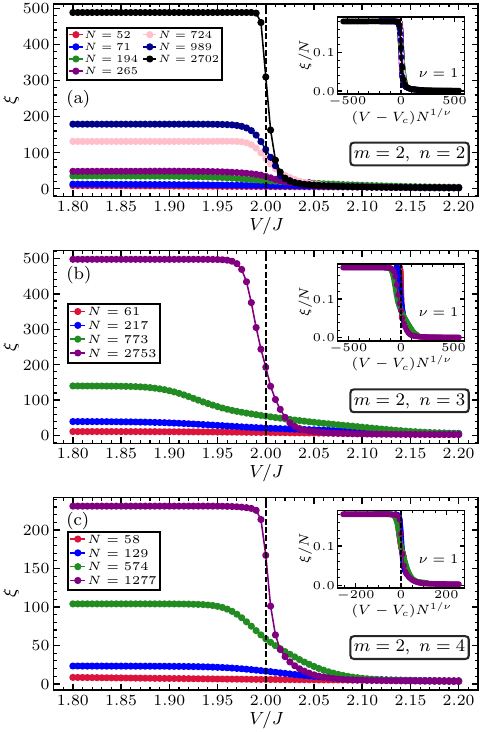}
\caption{
Localization length $\xi$ as a function of the quasiperiodic potential strength $V/J$ for $m=2$: 
(a) $n=2$; (b) $n=3$; (c) $n=4$. 
The insets show the rescaled localization length $\xi/N$ as a function of the scaling variable 
$N^{1/\nu}(V-V_c)$ for $\nu=1$. 
For all three cases, the critical point is fixed at the self-dual point $V_c=2J$, and the data collapse supports the correlation-length exponent $\nu=1$. 
All results shown here are averaged over 50 random samples of the phase $\phi$.
}
\label{fig:xi_m2_n234_collapse}
\end{figure}

\begin{figure}[htp]
    \centering
    \includegraphics[width=\columnwidth]{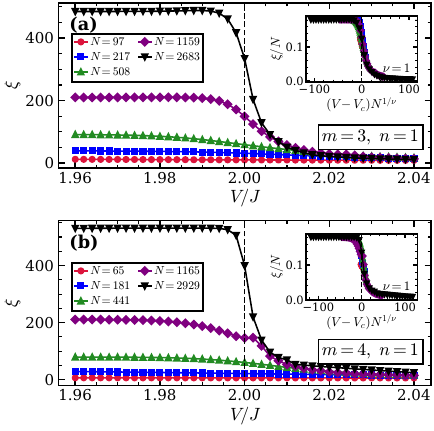}
    \caption{
    Localization length $\xi$ as a function of the quasiperiodic potential strength $V/J$ for $n=1$: 
    (a) $m=3$; (b) $m=4$. 
    The insets show the rescaled localization length $\xi/N$ as a function of the scaling variable $N^{1/\nu}(V-V_c)$ for $\nu=1$. 
    For all three cases, the critical point is fixed at the self-dual point $V_c/J=2$, and the data collapse supports the correlation-length exponent $\nu=1$. All results shown here are averaged over 50 random samples of the phase $\phi$. 
    }
    \label{fig:xi_n1_m345_collapse}
\end{figure}

We next examine the finite-size scaling behavior of the localization length for the other generalized Fibonacci modulations. Figure~\ref{fig:xi_m2_n234_collapse} shows the localization length $\xi$ as a function of the quasiperiodic potential strength $V/J$ for fixed $m=2$ and different values of $n$. The three cases are $(m,n)=(2,2)$, $(2,3)$, and $(2,4)$. For all cases, $\xi$ exhibits a clear change in behavior near the self-dual point $V_c/J=2$. On the extended side, the localization length increases with the system size, while on the localized side it remains finite. This behavior provides direct evidence for the localization transition at $V_c/J=2$. To further determine the correlation-length exponent, we perform a finite-size scaling collapse of the rescaled localization length $\xi/N$. According to the scaling form, $\xi/N$ should be a universal function of the scaling variable $N^{1/\nu}(V-V_c)$ near the critical point. As shown in the insets of Fig.~\ref{fig:xi_m2_n234_collapse}, the data for different system sizes collapse well when $V_c/J=2$ and $\nu=1$ are used. Therefore, the localization-length scaling for these $m=2$ generalized Fibonacci modulations is consistent with the correlation-length exponent $\nu=1$ in the AAH model.
We also study the localization-length scaling for the generalized Fibonacci sequences with $m>2$ and fixed $n=1$. Figure~\ref{fig:xi_n1_m345_collapse} shows $\xi$ as a function of $V/J$ for $(m,n)=(3,1)$, $(4,1)$, and $(5,1)$. 
The insets of Fig.~\ref{fig:xi_n1_m345_collapse} show the corresponding finite-size collapse of $\xi/N$ as a function of $N^{1/\nu}(V-V_c)$. Using the  value $\nu=1$, the curves for different system sizes collapse onto a single scaling function for $m=3$, and $4$. This result indicates that these $n=1$ sequences share the same correlation-length exponent as the standard AAH model.

For the cases with $m>1$, we did not obtain a satisfactory finite-size collapse when extracting the energy gap $\Delta$. Therefore, we were unable to reliably determine the critical exponent $z$. We speculate that this may be due to strong finite-size effects in the gap scaling for $m>1$. Larger system sizes may be required to approximate the irrational number more accurately using rational approximants, thereby achieving a better data collapse and enabling a more reliable extraction of the critical exponent $z$.

\section{Discussion and Summary}

\label{sec5}
The central result of this work is that the generalized Fibonacci irrational modulations leave the correlation-length exponent robustly invariant, while giving rise to a systematic tunability of the dynamical critical exponent. This observation reveals an unexpected separation between static and dynamical critical scaling in quasiperiodic localization transitions.

The robustness of the correlation-length exponent can be understood from the self-duality of the Aubry-Andr\'{e}-Harper (AAH) model. For all irrational modulations considered in this work, the Hamiltonian preserves the same dual transformation as the conventional AAH model. Consequently, the localization transition remains pinned at the self-dual point ($V_c=2$). The finite-size scaling analyses of both the fidelity susceptibility and the localization length consistently yield $\nu \simeq 1$, indicating that the static critical behavior belongs to the same universality class as the standard AAH transition. Therefore, changing the arithmetic structure of the irrational modulation does not modify the underlying localization fixed point governing the divergence of the correlation length.

In contrast, the dynamical critical exponent exhibits a pronounced dependence on the irrational modulation. For the $m=1$ family, both the generalized fidelity susceptibility and the finite-size gap scaling show that the dynamical exponent increases continuously with the arithmetic parameter $n$. The numerical results further indicate that the dependence on the irrational number is more naturally described in terms of $\Lambda_1=\frac{1}{\Omega}$, where $\Lambda_1$ is the largest eigenvalue of the generalized Fibonacci substitution matrix
$M=
\begin{pmatrix}
n & m \\
1 & 0
\end{pmatrix}$. 
Our data suggest an approximately linear relation between $z$ and $\Lambda_1$, whereas a logarithmic dependence on $\ln(1/\Omega)$ provides a less satisfactory description. This observation implies that the dynamical critical scaling is closely connected to the inflation hierarchy encoded in the quasiperiodic sequence.

A possible physical interpretation is that $\Lambda_1$ controls the separation of hierarchical length scales generated by successive inflation steps. As $\Lambda_1$ increases, the quasiperiodic structure develops increasingly separated characteristic scales. Since the critical spectrum of the AAH model is known to possess a hierarchical and self-similar structure, stronger inflation may enhance spectral fragmentation and accelerate the collapse of low-energy gaps. From the scaling relation $\Delta \sim N^{-z}$, a larger value of $z$ corresponds to a faster suppression of the characteristic energy scale with increasing system size. The observed growth of $z$ with $\Lambda_1$ is therefore consistent with the picture that the inflation hierarchy controls the low-energy dynamical scaling of the quasiperiodic critical point.

Although the numerical evidence strongly supports such a connection, the precise theoretical relation between the dynamical exponent and the arithmetic properties of the irrational modulation remains an open problem. In particular, it would be highly desirable to establish an analytical framework connecting the substitution hierarchy, spectral self-similarity, trace-map renormalization, or multifractal properties of critical eigenstates to the observed arithmetic dependence of $z$. Such a theory would provide a deeper understanding of how number-theoretic structures influence critical dynamics in deterministic aperiodic systems.

In summary, we have shown that generalized Fibonacci irrational modulations provide a simple route to tuning the dynamical critical exponent of quasiperiodic localization transitions without modifying the microscopic Hamiltonian. While the correlation-length exponent remains fixed at the AAH value $\nu \simeq 1$, the dynamical exponent can be continuously varied through the arithmetic structure of the irrational modulation. Our results suggest that irrational numbers are not merely technical ingredients used to generate quasiperiodicity, but may serve as physically relevant control parameters for dynamical criticality. This perspective opens a promising avenue toward arithmetic control of universality in deterministic quasiperiodic systems.

Importantly, the continuously tunable dynamical critical exponent $z$ predicted here should be directly accessible in existing experimental platforms, including ultracold atoms in optical lattices and photonic quasicrystals. In these systems, dynamical criticality is more naturally probed through quantum transport rather than direct measurements of finite-size energy gaps.

A standard protocol is to monitor the spreading of an initially localized wave packet. At criticality, the root-mean-square displacement typically follows an anomalous diffusion law, $\sqrt{\langle x^2(t)\rangle}\sim t^{\delta}$, where $\delta$ denotes the transport exponent. The transport dynamics is known to be strongly influenced by the multifractal structure of critical eigenstates and is governed by the same low-energy scaling properties that determine the dynamical critical exponent $z$. Consequently, one generally expects slower wave-packet expansion for larger values of $z$, corresponding approximately to $\delta\sim 1/z$.

Within this picture, the arithmetic tuning of $z$ should manifest itself as a systematic reduction of the transport exponent delta. Experimental observation of such a trend would provide direct evidence that the arithmetic structure of the underlying irrational modulation can control critical dynamics, thereby offering a practical route toward the experimental verification of arithmetic criticality in deterministic quasiperiodic systems.

\begin{acknowledgments}

The authors appreciate very helpful discussions with X. Wei. 
This work is supported by the National Natural Science Foundation of China (NSFC) under Grant Nos.~12404285, 12174194, the Zhejiang Provincial Natural Science Foundation of China under Grant No.~LQN25A040003, and the Science Foundation of Zhejiang Sci-Tech University under Grant No.~23062182-Y. 
W.-L. Y. acknowledges additional support
from the NUAA funding under Grant No. 1018-ILF26028.

\end{acknowledgments}
\section*{DATA AVAILABILITY}
The data that support the 
findings of this article are openly available \cite{Yi2026Github}.

\appendix

\section{Representative substitution sequences}
\label{app:substitution_sequences}

\begin{figure}[t]
\centering
\includegraphics[width=\linewidth]{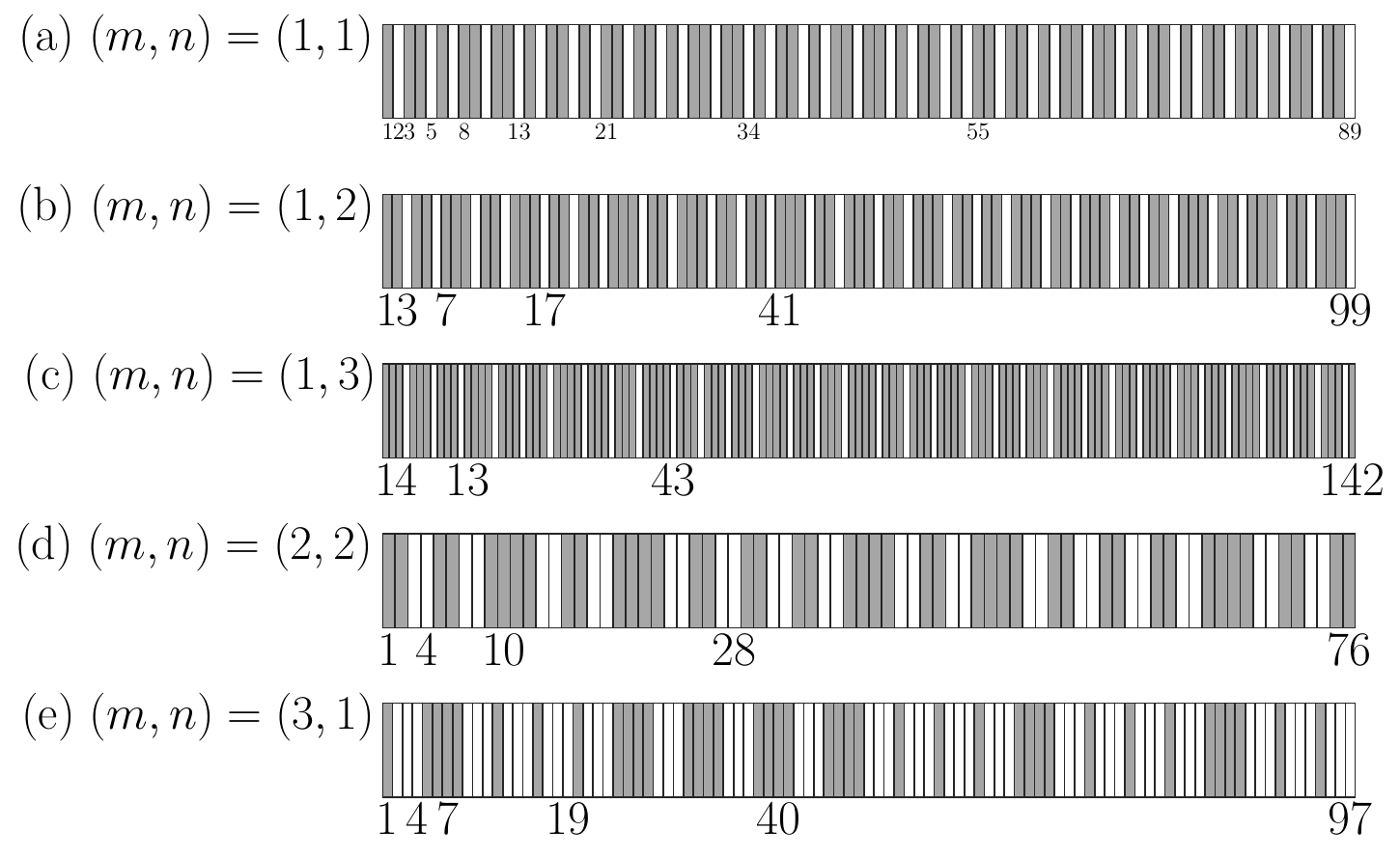}
\caption{
Schematic illustration of representative quasiperiodic sequences generated by the substitution rule 
$A \rightarrow A^{n}B^{m}$ and $B \rightarrow A$. 
Panels (a)–(e) correspond to $(m,n)=(1,1),(1,2),(1,3),(2,2)$, and $(3,1)$, respectively. 
The sequences are visualized as binary patterns, where different colors represent distinct local configurations. 
The integers shown below each panel indicate the system sizes $F_k$ used in the construction.
}
\label{cartoon}
\end{figure}
To illustrate the arithmetic structure underlying the generalized AAH model,
we consider the binary substitution rule
$A\rightarrow A^{n}B^{m}$ and $B\rightarrow A$. Starting from a single
letter $A$, repeated application of this rule generates a deterministic
hierarchical sequence composed of the two letters $A$ and $B$. The integers
$(m,n)$ control the relative abundance and arrangement of the two building
blocks, and hence define different generalized Fibonacci-type substitution
classes. In this sense, changing $(m,n)$ changes not only the rational
approximants to the irrational modulation frequency, but also the associated
inflation hierarchy.

Representative sequences generated in this way for several choices of
$(m,n)$ are shown in Fig.~\ref{cartoon}. These sequences are not introduced
as additional lattice potentials in the numerical calculations; rather, they
serve as a visual representation of the arithmetic hierarchy associated with
the corresponding irrational modulation. The figure makes clear that different
choices of $(m,n)$ lead to distinct binary patterns, reflecting the different
ways in which quasiperiodic order is built up through successive substitution
steps.

\bibliography{refs}

\end{document}